\documentclass[prd,aps,floats,floatfix,superscriptaddress,preprintnumbers,showpacs,eqsecnum,nofootinbib,twocolumn]{revtex4}

\usepackage{latexsym,array,theorem,mathrsfs,bm,float}
\usepackage{psfrag}
\usepackage{amsfonts,amsmath,amssymb,latexsym,array,afterpage,theorem,mathrsfs,bm,float,epsfig,color,graphicx,tabularx,here,multirow}
\usepackage[normalem]{ulem}

\newcommand{\gsim}{\, \mbox{\raisebox{-1.ex}
{$\stackrel{\textstyle>}{\textstyle\sim}$}}\,}
\newcommand{\lsim}{\, \mbox{\raisebox{-1.ex}
{$\stackrel{\textstyle<}{\textstyle\sim}$}}\,}

\newcommand{\bea}{\begin{eqnarray}}
\newcommand{\ena}{\end{eqnarray}}
\newcommand{\beann}{\begin{eqnarray*}}
\newcommand{\enann}{\end{eqnarray*}}

\newcommand{\dsl}{\pa \kern-0.5em /}
\newcommand{\la}{\lambda}
\newcommand{\pa}{\partial}

\newcommand{\nn}{\nonumber\\}

\newcommand{\vect}[1]{\!\!\!\mbox{\,~\boldmath $#1$}}

\newcommand{\mpl}{M_\mathrm{\rm P}}

\definecolor{red}{rgb}{1,0,0}
\definecolor{blue }{rgb}{0,0,1}
\definecolor{green}{rgb}{0,1,0}

\definecolor{mygreen}{rgb}{0,0.8,0}


\begin{document}

\baselineskip=12pt

\preprint{WU-AP/1907/19}

\title{Stability of Hybrid Higgs Inflation}

\author{\sc{Seiga Sato}}
\email{s.seiga"at"gravity.phys.waseda.ac.jp}
\affiliation{
Department of Physics, Waseda University,
Shinjuku, Tokyo 169-8555, Japan}
\author{\sc{Kei-ichi Maeda}}
\email{maeda"at"waseda.jp}
\affiliation{
Department of Physics, Waseda University,
Shinjuku, Tokyo 169-8555, Japan}
\affiliation{
 Waseda Institute for Advanced Study (WIAS),
Waseda University, Shinjuku, Tokyo 169-8050, Japan}

\date{\today}

\begin{abstract}
Hybrid Higgs inflation model is a hybrid type of the so-called Higgs inflation model with 
the conventional non-minimal $\xi$ coupling to a scalar curvature $R$ and new Higgs inflation model 
with the derivative coupling to the Einstein curvature tensor $G_{\mu\nu}$.
This model can explain all possible value of the tensor-to-scalar ratio $r$ 
 allowed by the present CMB constraints with an appropriate choice of coupling parameters.
However the derivative coupling may causes a gradient instability during oscillation phase after inflation just as in new Higgs inflation model.
Analyzing the behaviours of perturbations  during oscillation phase 
in the hybrid Higgs inflation model, 
we show that the unstable scalar modes found in new Higgs inflation model
are stabilized by  the non-minimal $\xi$ coupling 
for some range of the coupling parameters.
\end{abstract}

\maketitle



\section{Introduction} 
Big Bang cosmology is believed to be a standard  model of the Universe.
It is confirmed by the observations such as  the cosmological microwave background (CMB) and  the abundance of the light elements at the early time\cite{Aghanim:2018eyx}. 
Furthermore, an inflationary scenario \cite{Starobinsky:1980te,Guth:1980zm,Sato:1980yn,new_inflation1,new_inflation2,chaotic_inflation} is
now coming into  a standard model of the early stage of the universe. 
Not only it can solve many difficulties in  Big Bang cosmology  such as the horizon problem, but also 
it will provide a natural explanation of the origin of density perturbations.
Many inflation models have been  so far proposed and studied in detail\cite{Martin:2013tda}.

In most of inflation models, we assume  a scalar field, 
which is called an inflaton and 
is responsible for a rapid acceleration of the universe.
It  also  plays a  very important role in the late stage of the universe, because quantum fluctuations of the scalar field will provide the density perturbations of the Universe, which we are observing now.
However we do not have a natural candidate for an inflaton in particle physics yet.
In the standard model of particle physics, we have only one scalar field, i.e., the Higgs field.
In old inflation and new inflation models\cite{Guth:1980zm,Sato:1980yn,new_inflation1,new_inflation2}, we discuss inflationary scenario with Higgs field.
However it turns out that the density fluctuation is too large to explain the observed density perturbations\cite{Aghanim:2018eyx}.
Then some authors have 
tried to find a successful scenario with Higgs field by extending gravitational 
interactions from pure general relativistic one.
The Brans-Dicke type theory was discussed in the context of old and new inflationary scenario,
which are called the extended inflation\cite{extended_inflation} and soft inflation\cite{soft_inflation1,soft_inflation2}.
The gravitational non-minimal  couplings with Higgs field
($\xi \phi^2 R$) \footnote{ We shall call it $\xi$ coupling} or the curvature-squared model ($R^2$) have also studied\cite{spokoiny1984inflation,Maeda:1988rb,futamase1989chaotic}.

Then it was pointed out that if  $\xi\sim -10^4$, this model is highly
consistent with observations\cite{Higgs_inflation}.
Since then it is called the Higgs inflation.
In this paper we call it conventional Higgs inflation.
This model is quite similar to the $R^2$ inflation model proposed by 
Starobinsky\cite{Starobinsky:1980te}.
In fact we find the similarity between two models by use of a conformal transformation
\cite{Maeda:1988}.

In 2010, new type of Higgs inflation with a derivative coupling of a  Higgs field to the curvature
was proposed\cite{Germani:2010gm,Germani:2010ux,Germani:2014hqa},
which is called new Higgs inflation.
The derivative coupling is assumed to be the form of 
$G_{\mu\nu}\pa^\mu \phi\pa^\nu\phi/M^2$, where  $G_{\mu\nu}$ is the Einstein tensor and $M$ is a mass scale of the coupling.
Furthermore, generalized Higgs inflation\cite{kamada2012generalized} was proposed in 2014, which has all possible gravitational interactions with a single scalar field in the 
most general scalar tensor theory, i.e., the Horndeski theory.
Those ``non-minimal'' gravitational couplings may
 appear in the high energy limit such as an inflationary stage of the universe.
 Although those are most general models, we have to fix the arbitrary functions in the model
 in order to analyze the observational properties in detail.
 We have then proposed the hybid type of two Higgs inflation 
models (conventional and new Higgs inflation models), which we call 
hybrid Higgs inflation\cite{Sato:2017qau}.
We perform the disformal transformation and truncate the higher-derivative terms for our
analysis, which is justified with the slow-roll condition.
In this model, we show that 
 the primordial tilt $n_s$ barely changes,  while 
 the tensor-to-scalar ratio $r$ moves from the value in new Higgs inflationary model to that in the conventional Higgs inflationary model as $|\xi|$ increases.
Hence once we know the tensor-scalar ratio $r$ by the future observations, 
we can fix the coupling parameters $M$ and $\xi$. 

Although those Higgs inflation models are consistent with the CMB observations,
it is known that the derivative couplings in new Higgs inflation model 
cause gradient instabilities in the oscillation phase of the Higgs filed  
after inflation\cite{Ema:2015oaa}.
The sound speed of the perturbations becomes imaginary 
when the background Higgs field is oscillating.
It induces the exponential growth of the perturbations.
Even if the instability period is very short, high frequency modes 
grow large enough that the perturbation approach is no longer valid.

Since the hybrid Higgs inflation model has the derivative coupling, 
in this research, 
we study the perturbation dynamics  in the oscillation period
after inflation in the hybrid Higgs inflation and analyze its stability.
We show that the $\xi$ coupling term can stabilize 
 the instability induced by the derivative coupling.
We then present the stability constraints on 
the coupling parameters $\xi$ and $M$.

This paper is organized as fellows:
In \S. 2, we present our model and analyze the background dynamics 
in Jordan frame.
It is because we cannot truncate the higher-derivative terms induced by the disformal transformation  in the oscillation period after inflation.
In \S. 3, we write down the basic equations for perturbations in the oscillation period
and analyze stability against the perturbations.
We summarize  our results and give some remarks in \S. 4.
In this paper we use the units of  $\hbar=c=1$, but keep 
the reduced Planck mass $\mpl:=(8\pi G)^{-1/2}$ as it is.

\begin{widetext}
~~
\section{Hybrid Higgs Inflation :  Model and Background} 
The action of the hybrid Higgs inflation\cite{Sato:2017qau} is given by
\bea
S&=&\int d^4x \sqrt{-g}\left[ \frac{M_{\rm P}^2-\xi h ^2}{2} R
-\left( g^{\mu\nu}-\frac{G^{\mu\nu}}{M^2}\right) \frac{\partial_\mu h \partial_\nu h}{2}-V(h) \right],
\label{hha}
\\
&&
~~~~~~~~~~~~~~~{\rm with}~~~
V(h)=\frac{\la}{4}h^4
\,,
\nonumber
\ena
where $h$ is a Higgs field in a unitary gauge and $\lambda$ is its self-coupling constant. 
Here the potential $V$ is approximated in the high-energy limit of  inflationary stage
and the radiative correction is ignored just for simplicity.
This is a hybrid model of a conventional Higgs inflation and new Higgs inflation.

In our previous paper\cite{Sato:2017qau}, assuming a slow-roll condition
and performing a disformal transformation, 
we analyzed the inflationary stage.
Because of a slow-roll condition, we can ignore the higher-order derivative terms,
by which the analysis becomes much easier.
In fact, we obtain  an effective potential, which provides 
all information about the inflationary phase.
We find that some models with 
an appropriate choice of parameters $\xi$ and $M$ 
are consistent with observations.
The tensor-scalar ratio $r$
varies from the value in
the new Higgs inflationary model to that in the conventional Higgs inflationary model as
$|\xi|$ increases, while the spectral index  $n_s$ does not change so much 
for the parameters giving a right magnitude of density perturbations.

In this paper, since we are interested in the oscillation phase of the Higgs field, 
which is just after 
the inflationary stage, we cannot assume a slow-roll condition.
As a result, we have to analyze
very complicated equations with higher derivative terms in the Einstein frame
 or rather simpler equations in the original Jordan frame.
 We have chosen the latter case.
 
 As a background spacetime of the Universe, 
we adopt the flat Friedmann-Lema\^itre-Robertson-Walker (FLRW) spacetime, which metric is given by
\begin{equation*}
ds^2=- N^2(t) dt^2+a^2(t)d\vect{x}^2
\,,
\label{flrw}
\end{equation*}
where $N$ is a lapse function and $a$ is a scale factor of the Universe.
Assuming the background Higgs field also depend only on the cosmic time $t$,
 the action (\ref{hha}) for the background field is reduced to be
\begin{eqnarray}
S=\int d^4x\sqrt{-g}\Big{[} -3\left({H\over N}\right)^2\left(M_{\rm P}^2-\xi h^2\right)
+\left( 1+\frac{3H^2}{N^2 M^2}\right) \frac{\dot{h}^2}{2N^2}+6\xi {Hh\dot{h}\over N^2}
-V(h) \Big{]}\,, 
\end{eqnarray}
where a dot denotes a derivative with respect to  $t$ 
and $H\equiv\dot{a}/a$ is the Hubble expansion parameter.

Varying this action with respect to $h$, $N$ and $a$ and setting $N=1$, 
we obtain the basic equations for the background Universe as
\begin{align}
&
\tilde H^2=\frac{1}{3(1-\xi \tilde h^2)}\left[\frac{1}{2}\left(1+9\tilde H^2\right)\varpi_M^2
+6\xi \tilde h \tilde H\varpi_M+\tilde V\right],
\label{fr1} 
\\
&
{\cal M}
\left(
\begin{array}{c}
{\displaystyle {d\varpi\over d\tilde t}} \\[1em]
{\displaystyle {d\tilde H\over d \tilde t}}\\
\end{array}
\right)
=
-\left(
\begin{array}{c}
3\tilde H\varpi\left(1+3\tilde H^2\right)+12\xi \tilde h \tilde H^2+
{\displaystyle{d\tilde V\over d\tilde h}}\\
\varpi^2\left(1+3 \tilde H^2\right)+2\xi \varpi(\tilde H\tilde h-\varpi)\\
\end{array}
\right)
\label{eom}
\end{align}
with 
\begin{align}
{\cal M}\equiv 
\left(
\begin{array}{cc}
 1+3\tilde H^2&6\left(\tilde H\varpi+\xi \tilde h\right) \\
-2\left(\tilde H\varpi+\xi \tilde h\right)  & 2\left(1-\xi \tilde h^2-{\varpi^2\over 2}\right)\\
\end{array}
\right)
\end{align}
Here we have introduced dimension-free variables:
\beann
\tilde h={h\over M_p}\,,~~\varpi={\dot h \over MM_p}
\,,~~
\tilde H={H\over M}
\,,~~\tilde V={V\over M^2 M_p^2}
\,,~~
\tilde t=Mt
\enann
\end{widetext}

\noindent
Eq. (\rm{\ref{fr1}}) is the so-called Friedmann equation, which 
 gives a constraint on the background fields.
From Eq. (\rm{\ref{fr1}}), we find
\beann
&&
3\tilde H^2\left[1-\xi(1-6\xi)\tilde h^2-{3\over 2}\varpi^2\right]
\\
&&
={1\over 2}\left(
\varpi+6\xi \tilde H\tilde h\right)^2+\tilde V\,.
\enann
We then obtain the upper bound for $\varpi^2$ as
\bea
\varpi^2\leq {2\over 3}\left[1-\xi(1-6\xi)\tilde h^2\right]
\,.
\label{varpiupper}
\ena
Here we have assumed $\xi(1-6\xi)\leq 0$, i.e., $\xi\leq 0$ or $\xi\geq 1/6$.

A set of equations (\rm{\ref{eom}}) describes the equation of motion for the Higgs field $h$
and the equation of motion for the scale factor $a$, one of which is
derived from the other with the constraint (\rm{\ref{fr1}}).
In the equation for the Higgs field, the term with $\dot h$ 
gives a kind of viscosity caused by the cosmic expansion and 
it is modified due to the derivative coupling.
The $\xi$ coupling term behaves like an additional mass term.

Before going to  analyze the stability of our cosmological model against the perturbations,
we show the background behavior from inflationary stage to the oscillation phase
in Jordan frame.
For the background universe, we have to solve the above equations, i.e., Eq.  (\rm{\ref{fr1}})
and one of a set of equations (\rm{\ref{eom}}).

Two different behaviors of the evolution of Higgs field from the end of inflation to the oscillating 
phase are shown  in Fig.\rm{\ref{Fig:h_e_xiDom}}. 
Fig.\rm{\ref{Fig:h_e_xiDom}}(a) shows the evolution of the background Higgs field 
 when the $\xi$
coupling is dominant. We choose $\xi=-2.34\times10^{4}$ and $M=6\times10^{-4}M_P)$.
In this case, the Higgs field changes sharply near $\tilde{h}=0$, i.e., 
the time derivative of the Higgs field becomes quite large at the potential minimum.
We also present the evolution in the phase space,  in Fig. \rm{\ref{Fig:h_e_xiDom}}(b).
The arrows denote that the Higgs field evolves in that direction.
It shows a
very peculiar behavior in the Jordan frame, i.e., 
the orbit becomes very sharp near $h=0$ or $\varpi=0$.

On the other hand, when the derivative coupling is dominant, 
the Higgs field evolves more moderately near  $\tilde{h}=0$ as shown  in Fig. \rm{\ref{Fig:h_e_xiDom}}(c). We choose $\xi=-2.34\times10^{4}$ and $M=6\times10^{-4}M_P)$.
It shows a zigzag behavior during oscillations. 
We also show the evolution of the Higgs field in the phase
 space in Fig. \rm{\ref{Fig:h_e_xiDom}}(d).
The orbit becomes a deformed round shape, which is quite different from the previous 
$\xi$-coupling dominant case.

\begin{widetext}

\begin{figure}[h]
 \begin{center}
\includegraphics[width=5.5cm]{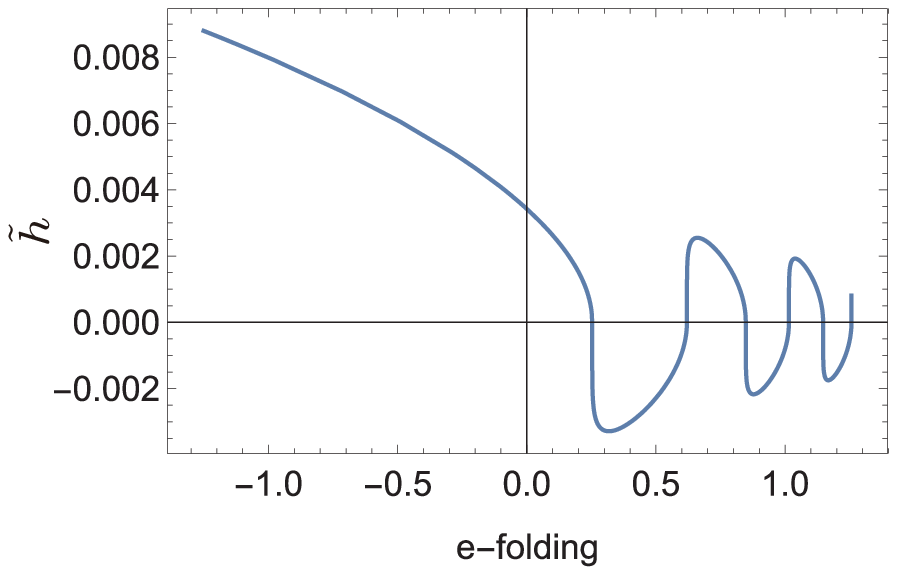}\hskip 2cm
\includegraphics[width=5.5cm]{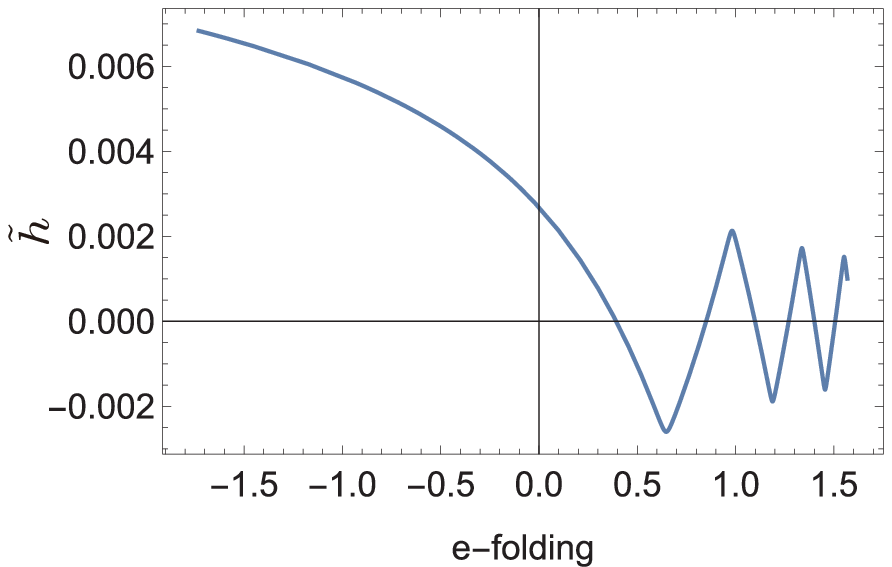}
\\
(a)\hskip 7cm (c)\\[1em]
\includegraphics[width=6cm]{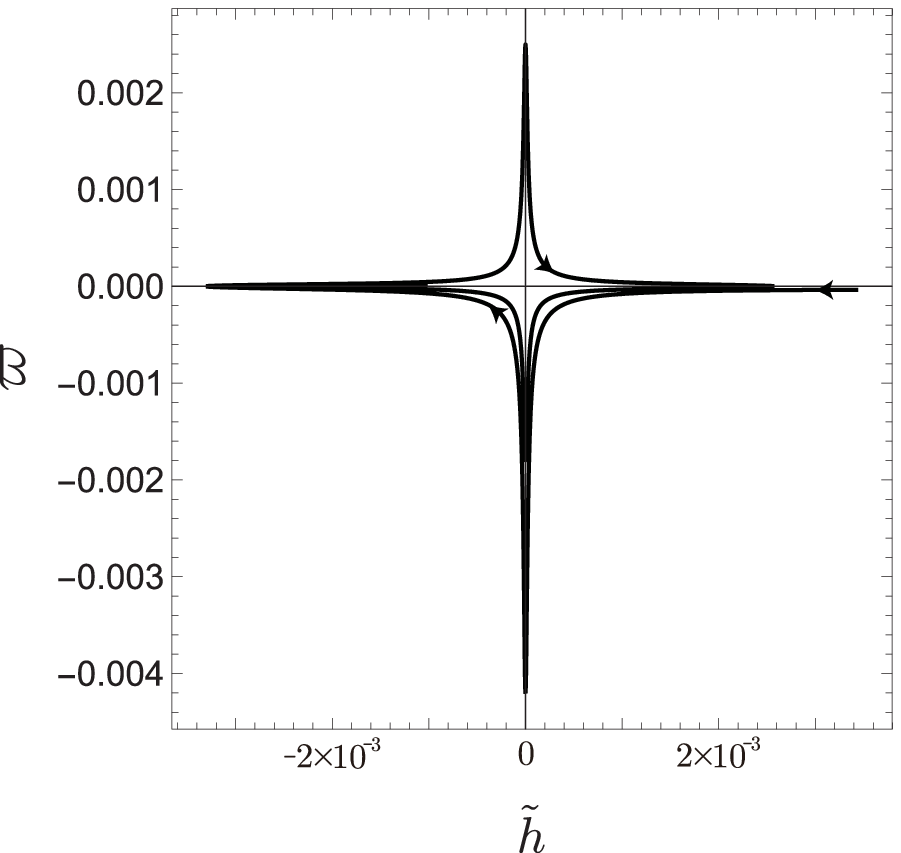}\hskip 2cm
 \includegraphics[width=5 cm]{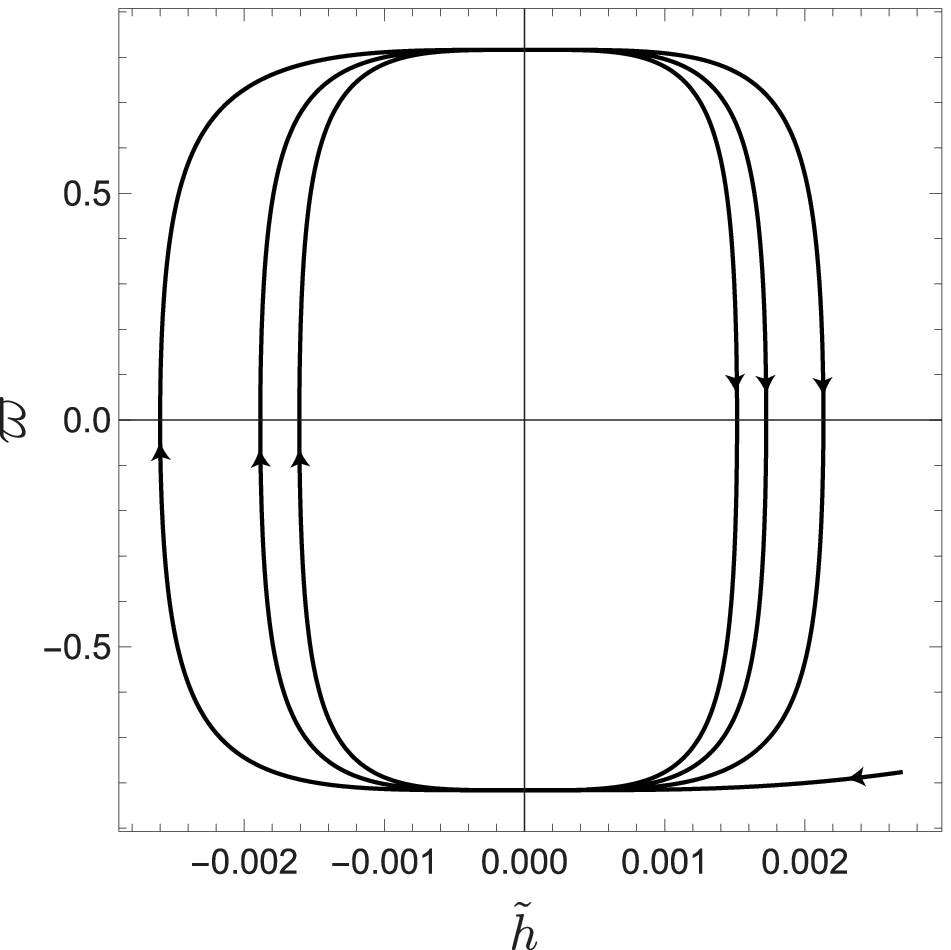}
\\
(b)\hskip 7cm (d)
 \end{center}
 \caption{The evolution of the Higgs field for the $\xi$-coupling dominant case
[(a) and (b)] and for
  the derivative-coupling dominant case ([(c) and (d)].
   (a)  and (c) show the time evolutions in terms of the $e$-folding, while (b) 
   and (d) gives
 the orbits in the  phase space.
 We set the $e$-folding $= 0$ at the end of the inflation.
 The orbits evolve along the arrows directions.
 We choose $ (\xi,M)=(-2.34\times10^{4},6\times10^{-4}M_P)$
 for the $\xi$-coupling dominant case
 and $ (\xi,M)=(-10,1.41\times10^{-8}M_P)$ for
  the derivative-coupling dominant case. 
}
 \label{Fig:h_e_xiDom}
\end{figure}

\section{Stability analysis} 
\label{SS:SoP}

\subsection{Perturbation Equations}
In order to investigate a stability of the present model,
 we perform perturbations of the background spacetime.
The perturbed metric is described by the ADM formalism as
\begin{equation}
ds^2 =-N^2 dt^2+\gamma_{ij}\left(dx^i+N^i dt\right)\left(dx^j+N^j dt\right),
\end{equation}
where the metric components are given by 
\begin{align}
N=1+\alpha ,~~~~N_i=\partial_i \beta ,~~~~
\gamma_{ij}=a^2(t)e^{2\zeta}\left(\delta_{ij}+h_{ij}+\frac{1}{2}h_{ik}h_{kj}\right)
\,.
\end{align}
$\alpha,\beta$, and $\zeta$  are the scalar modes, while  $h_{ij}$
describes the tensor modes.
We have ignored the vector modes, which may not play any crucial roles.

The tensor perturbation $h_{ij}$ 
satisfies the transverse-traceless conditions: $h_{ii}=\partial_j h_{ij}=0$.
Since we adopt the comoving gauge  in this paper, 
we do not need to take into account the perturbations of the Higgs field $h$.
\par
First, we give the equations for scalar perturbations.
The quadratic action for the scalar perturbations is found \cite{kobayashi2011generalized} as
\begin{equation}
S_S^{(2)}={M_{\rm P}^2\over 2 M^2}\int d\tilde t d^3\tilde x \, a^3\left[ G_S
(\tilde \partial_0\zeta)^2-\frac{F_S}{a^2}\left(\tilde \partial_i\zeta\right)^2\right]
\,,~~~
\label{qSS}
\end{equation}
where
\begin{eqnarray}
G_S&=& 
\frac{\varpi^2\left(1-\xi\tilde h^2-\frac{1}{2}\varpi^2\right)
\left[(1+3\tilde H^2)\left( 1-\xi \tilde h^2-{1\over 2}\varpi^2\right)
+6(\tilde H\varpi+\xi\tilde h)^2
\right]}{2\left[ \tilde H
\left(1 -\xi\tilde h^2-\frac{1}{2}\varpi^2\right)-\varpi(\tilde H\varpi+\xi\tilde h)\right]^2}
~~~~~~~~
\label{Gs}
\end{eqnarray}
\begin{eqnarray}
F_S
&=&
{\varpi^2\over (Hf-\varpi g)^2}\left[
{1+3H^2\over 2}f^2+(H\varpi+3\xi h)fg-\varpi^2 g^2+f(f\dot H+2g \dot \varpi)
\right]
\,,
\label{Fs}
\end{eqnarray}
with
\beann
f=1-\xi h^2-{\varpi^2\over 2}
\,,{\rm and}~~
g=H\varpi+\xi h
\,.
\enann
Here we have  replaced $\alpha$ and $\beta$ with $\zeta$ by use of
 two equations obtained from the constraint equations, which are given by  
\begin{eqnarray}
\alpha&=&\frac{1-\xi \tilde h^2-\frac{1}{2}\varpi^2}{\tilde H
\left(1 -\xi\tilde h^2-\frac{3}{2}\varpi^2\right)-\xi\tilde h\varpi}\tilde \partial_0 \zeta\,,
\\
\frac{\tilde \partial_i^2 \beta}{a^2}&=&-\frac{1-\xi \tilde h^2-\frac{1}{2}\varpi^2}{\tilde H
\left(1 -\xi\tilde h^2-\frac{3}{2}\varpi^2\right)-\xi\tilde h\varpi}\frac{\tilde \partial^2_i\zeta}{a^2}
\nn
&+&
{\varpi^2 \over 2}\Big{[}
3\tilde H^2(1-\xi\tilde h^2-{1\over 2}\varpi^2)+12\xi \tilde H\tilde h\varpi
+1-\xi(1-6\xi)\tilde h^2-{1\over 2}\varpi^2
\Big{]}
\tilde \partial_0 \zeta\,.
\end{eqnarray}
The sound speed $c_S$ is defined by
\begin{equation}
c_S^2=\frac{F_S}{G_S}
\,.
\end{equation}
For stability against perturbations such that
there exist
no tachionic instability and no gradient instability, 
we have to impose 
\bea
F_S\geq 0
~~{\rm and}~~~
c_S^2\geq 0
\,.
\label{stability_cond_scalar}
\ena

Similarly  we obtain the quadratic action for the tensor perturbations as
\bea
S_T^{(2)}&=&{M_{\rm P}^2\over 2M^2}\int d\tilde td^3\tilde x \, a^3
\Big{[} G_T\left(\tilde \partial_0 {h}_{ij}\right)^2
-\frac{F_T}{a^2}\left(\tilde \partial_k  h_{ij}\right)^2\Big{]},
\label{qST}
\ena
where $G_T$ and $F_T$ are
\begin{align}
G_T=1 -\xi\tilde h^2-\frac{1}{2}\varpi^2\,,
{\rm and}~~~
F_T=1 -\xi\tilde h^2+\frac{1}{2}\varpi^2\,.
\end{align}
\end{widetext}
We then find the stability conditions as
\bea
F_T\geq 0
~~{\rm and}~~~
c_T^2={F_T\over G_T}\geq 0
\,.
\label{stability_cond_tensor}
\ena

\subsection{Stability of conventional and new Higgs inflation models}
When we have only $\xi$ coupling term, 
we can show that
the background spacetime is stable against perturbations, i.e., 
the stability conditions 
(\ref{stability_cond_scalar}) and (\ref{stability_cond_tensor}) are always satisfied.
It is naturally expected since the system described in the Einstein frame
 is given by the Einstein gravity plus a scalar 
field with a positive definite potential. 

\begin{figure}[thb]
 \begin{center}
 \includegraphics[width=6cm]{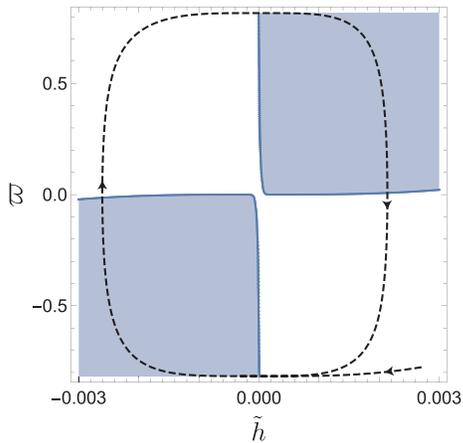}
 \end{center}
 \caption{The phase space of the Higgs field  for new Higgs inflation. The dotted line represents the evolution of the Higgs field of the first oscillation just after the end of inflation, which path is coming from the bottom-right region and evolves along with arrows. The $F_S$ becomes negative in the shaded region. Since the dotted line goes through the shaded region, the gradient instability occurs.}
 \label{Fig:InstaPhasexi0}
\end{figure}

For new Higgs inflation ($\xi=0$), both $F_T$ and $G_T$ are always positive.
The tensor perturbations do not cause any instability.
However, the scalar perturbations show unstable behaviors in the oscillation phase.
Although $G_S$ is always positive 
 due to the (\rm{\ref{varpiupper}}), $F_S$ becomes negative during the oscillations.
One concrete example is given in Fig.\ref{Fig:InstaPhasexi0}, in which 
we present the negative $F_S$-region shown by the shaded color in the phase space of the Higgs field.
We also show the trajectory of the Higgs field just after the end of inflation by the dotted line.

Since the trajectory evolves into the negative $F_S$-region, 
the gradient instability occurs in such a region. 
The perturbations grows exponentially and becomes larger than the background value
soon.
This instability is always found in new Higgs inflation model ($M\neq  ~ \infty, \xi=0$).
As a result, new Higgs inflation model does not provide the reheating process after inflation.

\subsection{Stability of hybrid Higgs inflation model}
In the case of hybrid Higgs inflation model, 
we have to check whether the instability found in new Higgs inflation still exists or not, and 
if Yes, how the instability depends on the coupling parameters $\xi$ and $M$.
It is because the conventional Higgs inflation ($M=\infty, \xi\neq 0$) has no instability
and then $\xi$ coupling term ($\xi\neq 0$) may stabilize the system.
We then survey stability of the models with the parameters satisfying the observational constraint of 
the density perturbation ${\cal P}_\zeta$, which gives some relation between $M$ and $\xi$
\cite{Sato:2017qau}.

\begin{figure}[thb]
 \begin{center}
 \includegraphics[width=7.5cm]{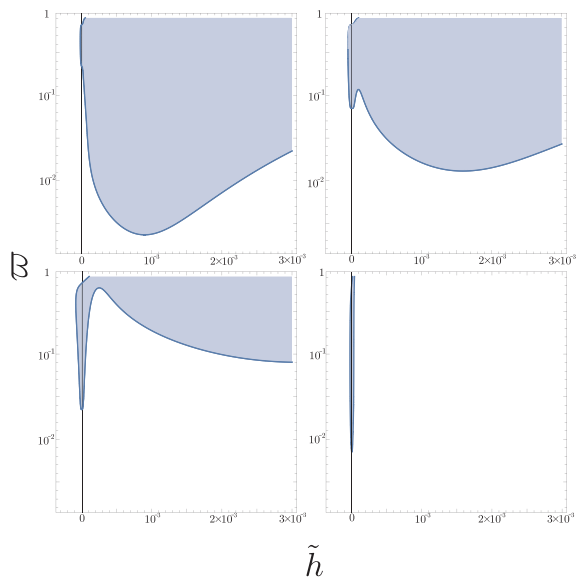}
 \end{center}
 \caption{The  negative $F_S$ region in the phase space of the Higgs field. 
We show only the region of $\varpi\geq0$.
The parameters in each figure are chosen as 
$(\xi,M)=(-10,1.41\times 10^{-8}M_P)$ [upper left], 
$(-10^2,1.41\times 10^{-8}M_P)$ [upper right], 
$(-10^3,1.70\times 10^{-7}M_P)$ [lower left], and 
$(-10^4, 3.98\times 10^{-8}M_P)$ [lower right], which
 satisfy the observational constraint on ${\cal P}_\zeta$. 
 As $|\xi|$ becomes larger, the unstable region is squeezed into 
 very narrow area along $\tilde{h}=0$.}
\label{Fig:Fssign_inM_fixs}
\end{figure}

The unstable region shown in Fig.\ref{Fig:InstaPhasexi0} for $\xi=0$ is 
 deformed as the $|\xi|$ becomes larger, as shown in Fig.\ref{Fig:Fssign_inM_fixs}.
Since $F_S$ is invariant under the transformation of $(\tilde{h},\varpi)\rightarrow(-\tilde{h},-\varpi)$, we plot only the region of $\varpi\geq 0$.

We then study the evolution of the Higgs field after inflation.
 Although the unstable region becomes much smaller as $|\xi|$ gets large, 
 the analysis shows that the orbits of the Higgs field in the phase space 
 always go through the unstable regions showed in Fig.\ref{Fig:Fssign_inM_fixs}
 unless $|\xi|$ is extremely large.
In fact we do not find a stable model for $|\xi|\leq 10^4$.

\begin{figure}[h]
 \begin{center}
 \includegraphics[width=7cm]{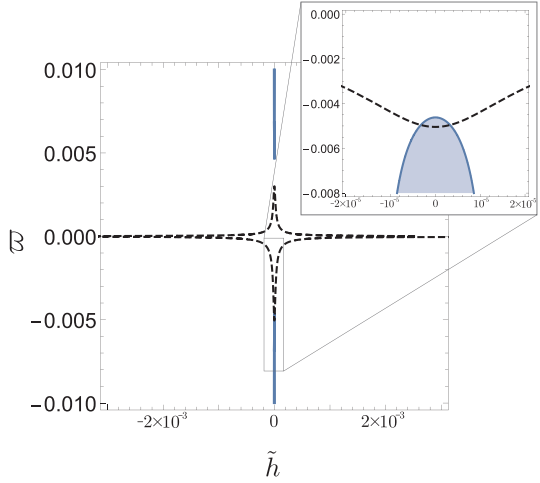}
 \end{center}
 \caption{The unstable region  in the phase space of the Higgs field
 and its orbit just after the end of 
  inflation. The parameters are chosen as $(\xi,M)=(-2.34\times 10^4, 5\times 10^{-4}M_P)$. 
  The dotted line shows the orbit of the Higgs field, while
   the blue shaded regions depicts the negative $F_S$ region. 
   The Higgs field evolves into the unstable region. We also enlarge the crucial part.}
\label{Fig:criticalout}
\end{figure}

\begin{figure}[h]
 \begin{center}
 \includegraphics[width=7cm]{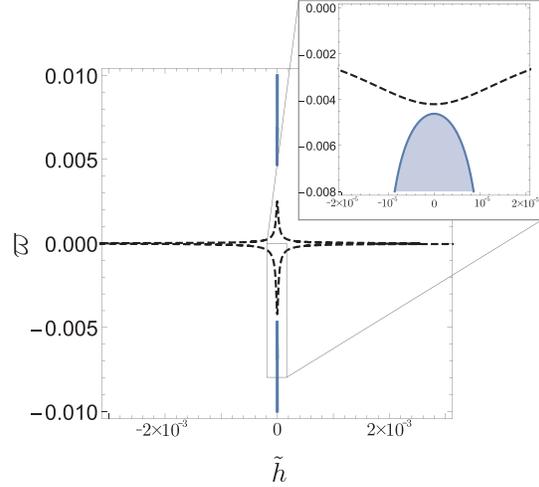}
 \end{center}
 \caption{The same figure  as Fig. \ref{Fig:criticalout}, but with the 
 different values of the parameters. $(\xi,M)=(-2.34\times 10^4, 6\times 10^{-4}M_P)$.
   The orbit of the Higgs field does not cross the unstable region. Hence this model is stable.}
\label{Fig:criticalok}
\end{figure}

For the case with $|\xi|>10^4$, the unstable region is squeezed into 
 very narrow area along $\tilde{h}=0$ just as the lower right figure in Fig.\ref{Fig:Fssign_inM_fixs}.
 Hence we can focus our analysis near the region of $h=0$.
 We find that the minimum value of $|\varpi|$ at $\tilde{h}=0$ in the negative $F_S$ region 
 is proportional to the $M^{-1}$, and 
 there exists a critical value of $M$, beyond which 
 the system is stable, i.e., the orbit of the Higgs field does not cross 
 the unstable region.
 We show the critical value should be between $M=5\times 10^{-4}M_P$, which 
 gives an unstable model (see Fig.\ref{Fig:criticalout}), and $6\times 10^{-4}M_P$, which
 gives a stable model  (see Fig.\ref{Fig:criticalok}).
For those values of $M$, 
$\xi=-2.34\times 10^4$ does hardly move because of  the observational constraint on ${\cal P}_\zeta$. \\[1em]

\vskip 1cm
We shall confirm this fact by semi-analytic approach.
Since the unstable region is localized  the area near $h=0$ for large $|\xi|$, it will be enough to 
analyze whether  the stable condition $F_S\geq 0$ is satisfied 
or not when the Higgs field passes through the point $h=0$.

In oscillating phase, $\varpi^2$ takes the maximum value near $\tilde{h}=0$.
In particular we consider the maximum value  $\varpi^2=\varpi_{\rm max}^2$ in
 the first oscillation cycle just after the end of  inflation.
It is because since the oscillation is dissipative because of the expansion of the universe,
 $\varpi^2$ near $\tilde{h}=0$ decreases in the evolution of the universe.\\
 
 As a result, we find $\varpi^2\leq \varpi_{\rm max}^2$ at  $\tilde{h}=0$.
 Hence if  the point of $\varpi_{\rm max}^2$ near $\tilde{h}=0$ is included in the stable region of $F_S\geq 0$ in the phase space,
 the system is stable.

\begin{widetext}
Eqs.  (\rm{\ref{Fs}}) and (\rm{\ref{Gs}}) with $\varpi_{\rm max}^2$ and  $\tilde{h}=0$ 
become
\begin{align}
F_S=&\frac{12\left( 1-\frac{3}{2}\varpi_{\rm max}^2\right)^2 -8\varpi_{\rm max}^4\left( 1-\frac{3}{2}\varpi_{\rm max}^2 \right)-3\varpi_{\rm max}^8+24\xi\varpi_{\rm max}^2\left( 1-\frac{\varpi_{\rm max}^2}{2}\right)\left( 1-\frac{3}{2}\varpi_{\rm max}^2-\frac{3}{2}\varpi_{\rm max}^4\right)}{4\left( 1-\frac{3}{2}\varpi_{\rm max}^2\right)\left( 1-\frac{3}{2}\varpi_{\rm max}^2+\frac{3}{2}\varpi_{\rm max}^4\right)}
\,,
\label{Fs2}\\
G_S=&\frac{3\left( 1-\frac{\varpi_{\rm max}^2}{2}\right)\left( 1-\frac{3}{2}\varpi_{\rm max}^2+\frac{3}{2}\varpi_{\rm max}^4\right)}{\left( 1-\frac{3}{2}\varpi_{\rm max}^2\right)^2}
\,.
\end{align}
$G_S$ and the denominators of $F_S$ are always positive as long as $\varpi_{\rm max}$ 
keeps in the range of (\rm{\ref{varpiupper}}).
Hence, when the numerator of Eq.(\rm{\ref{Fs2}}) is positive, 
the system is stable. 
The stability condition for $\varpi_{\rm max}$ is then  given by
\begin{align}
\xi\leq -\frac{12\left( 1-\frac{3}{2}\varpi_{\rm max}^2\right)^2 -8 \varpi_{\rm max}^4\left( 1-\frac{3}{2}\varpi_{\rm max}^2\right)-3 \varpi_{\rm max}^8}{24\varpi_{\rm max}^2\left( 1-\frac{1}{2}\varpi_{\rm max}^2\right)\left( 1-\frac{3}{2}\varpi_{\rm max}^2-\frac{3}{2}\varpi_{\rm max}^4\right)} 
\end{align}
\end{widetext}

\begin{figure}[h]
 \begin{center}
 \includegraphics[width=7cm]{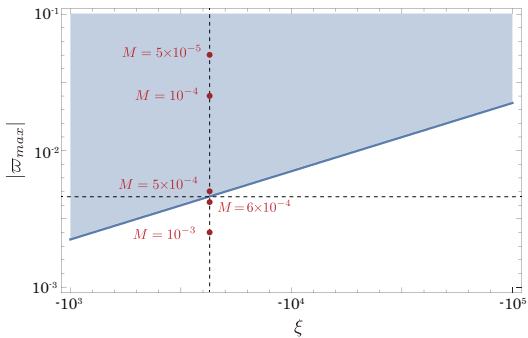}
 \end{center}
 \caption{The unstable condition is shown by the blue-shaded region.
  If the value of $\left|\varpi_{\rm max}\right|$, which is the maximum value of $|\varpi|$
in the first oscillation,  is too large,
 the orbit of the Higgs field will get into the blue-shaded unstable region. The dotted lines
 denote  $\xi\simeq-2.34\times 10^4$ and $\left|\varpi_{\rm max}\right|\simeq4.62\times 10^{-3}$.
The red dots denote the value $\left|\varpi\right|$ at $\tilde{h}=0$ for each $M$, which are the almost same as $\left|\varpi_{\rm max}\right|$.
}
 \label{Fig:Fsh0stable}
\end{figure}

This equation is very complicated as the equation for $\varpi_{\rm max}$ with a given value of $\xi$. However it turns out that it is rather simple for the range of  $\xi<-10^3$, which we are interested in.
We show this condition in this range in Fig.\rm{\ref{Fig:Fsh0stable}}.

The border of stability looks very simple. In fact  we can approximate 
the stability condition as
\beann
\log_{10}[|\varpi_{\rm max}|]\lsim -2.34-\frac{1}{2}(\log_{10}[-\xi] +4.37)
\,.
\enann
For given values of $\xi$ and $M$, we can evaluate $\varpi_{\rm max}$, which is
also shown by red dots in Fig. \ref{Fig:Fsh0stable}.
We find the critical value $M_{\rm cr}$ exists between $5\times 10^{-4}M_{\rm P}$ and $6\times 10^{-4}M_{\rm P}$.
We then obtain the condition for the parameters for stable hybrid Higgs inflation model
as follows:
\beann
\xi\approx -2.34\times 10^4\,,~~M\leq M_{\rm cr}\approx 6\times 10^{-4} M_{\rm P}
\,.
\enann

We conclude that there exists a stable hybrid Higgs inflation model, but the derivative coupling must be enough small enough.
The critical value for stability is 
$M_{\rm cr}\approx 6\times 10^{-4} M_{\rm P}$, 
below which the model becomes unstable against perturbations.
From the observational constraint on the density perturbation ${\cal P}_\zeta$, 
the coupling constant $|\xi|$ must be very large just as the conventional Higgs inflation model.
If the derivative coupling is large (corresponding to small $M$), 
the instability occurs in the oscillation phase just after 
the end of inflation (see Fig.\ref{Fig:Fssign_inM_fixs}).

\section{Concluding Remarks} 
We have analyzed the dynamics of the hybrid Higgs inflation in the oscillation phase just after inflation. The model contains two coupling of the Higgs field with the spacetime curvature;
 $\xi$ coupling and  the derivative coupling.
The derivative coupling may sometimes cause the gradient instability.
In fact in new Higgs inflation model, the scalar modes of the perturbations exponentially grow while the background Higgs field oscillates.
In such a model, we may not discuss a history of the universe after inflation.

In this paper, we have shown
 that this type of instability can be avoided when the $\xi$ coupling is dominant.
If the derivative coupling constant $M$ is smaller than  $M_{\rm cr}\approx 6\times 10^{-4}
M_{\rm P}$, the Higgs field evolves into the unstable region.
(see Figs. \ref{Fig:criticalout} and \ref{Fig:criticalok}).
With the observational constraint on the density perturbations ${\cal P}_\zeta$, we find the constraints on the parameters as $\xi\approx -2.34\times10{-4}$ and $M\gsim M_{\rm cr}$
for stable hybrid Higgs inflation model.

We should mention about the quantum effect on the Higgs field.
Although we have studied the stability of the Higgs inflation model during the 
oscillation phase at the tree level here, 
we should remind that the Higgs potential is very sensitive to the quantum loop effects\cite{Isidori:2001bm,Bezrukov:2009db,EliasMiro:2011aa,Degrassi:2012ry}.
The gravitational couplings which we discussed in this paper are 
assumed to appear due to the quantum gravity effects.
Hence it is natural to include the quantum loop effects on the Higgs field potential too.
However, the calculation of quantum loop effects is still under discussion\cite{bezrukov2015living,hamada2015higgs,Shaposhnikov:2018nnm}.
We leave it for the future works.

Since the gradient instability is avoided when the $\xi$ coupling is included,
we can analyze the reheating process after inflation now.
We expect that the similar analysis to the case in
 the the conventional Higgs inflation can be performed.
The work on the reheating process in the hybrid Higgs inflation model is in progress, 
and  the results will be published elsewhere.

\section*{Acknowledgements}
This work was supported in part by JSPS KAKENHI Grant Numbers   
JP17H06359 and JP19K03857, and by Waseda University Grant for Special
Research Projects (Project number: 2019C-254 and 2019C-562). 


\end{document}